\documentstyle[epsf,psfig,12pt]{article}

\renewcommand{\baselinestretch}{1.10}

\begin{document}

\renewcommand{\theequation}{\thesection.\arabic{equation}}
\def\be{\begin{eqnarray}}
\def\en{\end{eqnarray}}
\def\non{\nonumber}
\def\la{\langle}
\def\ra{\rangle}
\def\up{\uparrow}
\def\down{\downarrow}
\def\ep{\varepsilon}
\def\ums{{\mu}_{_{\overline{\rm MS}}}}
\def\MS{{\overline{\rm MS}}}
\def\u{\mu_{\rm fact}}
\def\gg{\Delta\sigma^{\gamma G}}
\def\lsim{ {\ \lower-1.2pt\vbox{\hbox{\rlap{$<$}\lower5pt\vbox{\hbox{$\sim$}
}}}\ } }
\def\gsim{ {\ \lower-1.2pt\vbox{\hbox{\rlap{$>$}\lower5pt\vbox{\hbox{$\sim$}
}}}\ } }
\def\dk{\partial\!\cdot\!K}\def\dk{\partial\!\cdot\!K}
\def\pr{{\sl Phys. Rev.}~}
\def\prl{{\sl Phys. Rev. Lett.}~}
\def\pl{{\sl Phys. Lett.}~}
\def\np{{\sl Nucl. Phys.}~}
\def\zp{{\sl Z. Phys.}~}

\title{
\vskip 10mm \Large\bf  The Proton Spin Puzzle: A Status
Report$^{*}$ }
\author{{\bf Hai-Yang Cheng}\\
{\it Institute of Physics, Academia Sinica}\\ {\it Taipei, Taiwan
115, Republic of China}\\ }

\maketitle

\begin{abstract}
The proton spin puzzle inspired by the EMC experiment and its
present status are closely examined. Recent experimental progress
is reviewed. Various factorization schemes due to the ambiguity
arising from the axial anomaly are discussed. Some misconceptions
in the literature about the $\MS$ factorization scheme are
clarified. It is stressed that the polarized nucleon structure
function $g_1(x)$ is independent of the factorization scheme
chosen in defining the quark spin density. Consequently, the
anomalous gluon and sea-quark interpretations for the deviation of
the observed first moment of $g_1^p(x)$ from the Ellis-Jaffe sum
rule are equivalent. While it is well known that the total quark
spin in the chiral-invariant (CI) factorization scheme (e.g. the
improved parton model) can be made to be close to the quark model
expectation provided that the gluon spin is positive and large
enough, it is much less known that, contrary to the
gauge-invariant scheme (e.g. the $\MS$ scheme), the quark orbital
angular momentum in the CI scheme deviates even farther from the
relativistic quark model prediction. Recent developments in the
NLO analysis of polarized DIS data, orbital angular momentum and
lattice calculations of the proton spin content are briefly
sketched.

\vskip 4 cm \centerline{ $^*$ Invited plenary talk presented at
the} \centerline{ {\it 2000 Annual Meeting of the Physical Society
of ROC,}} \centerline{National Cheng-Kung University, Taiwan, ROC}
\centerline{January 31 - February 1, 2000 }
\end{abstract}
\pagebreak

\section{Introduction}
Experimentally, the polarized structure functions $g_1$ and $g_2$
are determined by measuring two asymmetries:
\be
A_\Vert=\,{d\sigma^{\up\down}-d\sigma^{\up\up}\over
d\sigma^{\up\down}+
d\sigma^{\up\up}},~~~~~A_\perp=\,{d\sigma^{\down\to}-d\sigma^{\up\to}\over
d\sigma^{\down\to}+d\sigma^{\up\to}},
\en
where $d\sigma^{\up\up}$ ($d\sigma^{\up\down}$) is the
differential cross section for the longitudinal lepton spin
parallel (antiparallel) to the longitudinal nucleon spin, and
$d\sigma^{\down\to}$ ($d\sigma^{\up\to}$) is the differential
cross section for the lepton spin antiparallel (parallel) to the
lepton momentum and nucleon spin direction transverse to the
lepton momentum and towards the direction of the scattered lepton.
From the parton-model or from the OPE approach, the first moment
of the polarized proton structure function
\be
\Gamma_1^p(Q^2) \equiv\int^1_0 g_1^p(x,Q^2)dx,
\en
can be related to the combinations of the quark spin components
via
\be
\Gamma_1^p=\,{1\over 2}\sum_q e^2_q\Delta q(Q^2)={1\over 2}\sum_q
e^2_q \la p,s|\bar{q}\gamma_\mu\gamma_5q|p,s\ra s^\mu,
\en
where $\Delta q$ represents the net helicity of the quark flavor
$q$ along the direction of the proton spin in the infinite
momentum frame:
\be
\Delta q=\int^1_0\Delta q(x)dx\equiv\int^1_0\left[
q^\up(x)+\bar{q}^\up(x)- q^\down(x)-\bar{q}^\down(x)\right]dx.
\en
At energies $\la Q^2\ra\sim 10\,{\rm GeV}^2$ or smaller, only
three light flavors are relevant:
\be \label{Gamma1}
\Gamma_1^p(Q^2)=\,{1\over 2}\left({4\over 9}\Delta u(Q^2)+{1\over
9}\Delta d(Q^2)+{1\over 9}\Delta s(Q^2)\right).
\en
Other information on the quark polarization is available from the
low-energy nucleon axial coupling constants $g_A^3$ and $g_A^8$:
\be
\label{gA38} g_A^3(Q^2)&\equiv &\la p,s|\bar u\gamma_\mu\gamma_5
u-\bar d\gamma_\mu\gamma_5 d|p,s\ra s^\mu=\Delta u(Q^2)-\Delta
d(Q^2),
\\ g_A^8(Q^2)&\equiv &\la p,s|\bar u\gamma_\mu\gamma_5 u+\bar
d\gamma_\mu\gamma_5 d-2\bar s\gamma_\mu\gamma_5 s|p,s\ra
s^\mu=\Delta u(Q^2)+\Delta d (Q^2)-2\Delta s(Q^2). \non
\en
Since there is no anomalous dimension associated with the
axial-vector currents $A_\mu^3$ and $A_\mu^8$, the non-singlet
couplings $g_A^3$ and $g_A^8$ do not evolve with $Q^2$ and hence
can be determined at $q^2=0$ from low-energy neutron and hyperon
beta decays. Under SU(3)-flavor symmetry, the non-singlet
couplings are related to the SU(3) parameters $F$ and $D$ by
\be
g_A^3=\,F+D,~~~~~~g_A^8=\,3F-D.
\en
We use the updated coupling $g_A^3=1.2670\pm 0.0035$ \cite{PDG}
and the values \cite{Goto}
\be
F=0.463\pm 0.008\,,~~~~D=\,0.804\pm 0.008\,,~~~~F/D=\,0.576\pm
0.016
\en
to obtain $g_A^8=0.585\pm 0.025$\,.

Prior to the EMC measurement of polarized structure functions, a
prediction for $\Gamma_1^p$ was made based on the assumption that
the strange sea in the nucleon is unpolarized, i.e., $\Delta s=0$.
It follows from (\ref{Gamma1}) and (\ref{gA38}) that
\be
\Gamma_1^p(Q^2)=\,{1\over 12}g_A^3+{5\over 36}g_A^8.
\en
This is the Ellis-Jaffe sum rule \cite{EJ}: $\Gamma_1^p=0.185\pm
0.003$ in the absence of QCD corrections and equals to $0.171\pm
0.006$ at $Q^2=10\,{\rm GeV}^2$ to leading-order corrections. The
1987 EMC experiment \cite{EMC} then came to a surprise. The result
published in 1988 and later indicated that $\Gamma_1^p=0.126\pm
0.018$, substantially lower than the expectation from the
Ellis-Jaffe conjecture.  From the EMC measurement of $\Gamma_1^p$,
we obtain
\be
\Delta u=\,0.77\pm 0.06\,,~~~\Delta d=-0.49\pm 0.06\,,~~~\Delta
s=-0.15\pm 0.06\,,
\en
and
\be
g_A^0(Q^2)&\equiv &\la p,s|\bar u\gamma_\mu\gamma_5 u+\bar
d\gamma_\mu\gamma_5 d+\bar s\gamma_\mu\gamma_5 s|p,s\ra s^\mu \non
\\
&=& \Delta u(Q^2)+\Delta d(Q^2)+\Delta s(Q^2)\equiv
\Delta\Sigma(Q^2)=\,0.14\pm 0.18
\en
at $Q^2=10.7\,{\rm GeV}^2$. The EMC results exhibit two surprising
features: The strange-sea polarization is sizeable and negative,
and the total contribution of quark helicities to the proton spin
is small and consistent with zero.  This is sometimes referred to
as the ``proton spin crisis".

The so-called ``proton spin crisis" is not pertinent since the
proton helicity content explored in the DIS experiment is,
strictly speaking, defined in the infinite momentum frame in terms
of QCD current quarks and gluons, whereas the spin structure of
the proton in the proton rest frame is referred to the constituent
quarks. That is, the quark helicity $\Delta q$ defined in the
infinite momentum frame is generally not the same as the
constituent quark spin component in the proton rest frame, just
like that it is not sensible to compare apple with orange. What
trigged by the EMC experiment is the ``proton helicity
decomposition puzzle" rather than the ``proton spin crisis" (for a
review, see
\cite{Anselmino,Jaffe96a,EK96,Cheng96b,Lampe,Shore98,Bass99}). The
non-relativistic SU(6) constituent quark model predicts
$\Delta\Sigma'\equiv\Delta U+\Delta D=1$ ($\Delta U$ denoting the
constituent up quark spin and likewise for $\Delta D$), but its
prediction for the axial-vector coupling constant $g_A^3={5\over
3}$ is too large compared to the measured value of $1.2670\pm
0.0035$ \cite{PDG}. In the relativistic quark model, the proton is
no longer a low-lying $S$-wave state since the quark orbital
angular momentum is nonvanishing due to the presence of quark
transverse momentum in the lower component of the Dirac spinor.
Realistic models e.g. the cloudy bag model \cite{Schreiber},
predict $\Delta\Sigma'\approx 0.60$; that is, about 40\% of the
proton spin is carried by the orbital angular momentum of the
constituent quarks. In the parton picture, the naive expectation
of $\Delta\Sigma$, which is equal to $g_A^8=0.59$ under the
assumption of vanishing sea polarization, is very close to the
relativistic quark model's prediction of $\Delta\Sigma'$. One of
the main theoretical problems is that hard gluons cannot induce
sea polarization perturbatively for massless quarks due to
helicity conservation. Hence, it is difficult to accommodate a
large strange-sea polarization in the naive parton model.

\section{Experimental Progress}
\setcounter{equation}{0}
Before 1993 it took 5 years on the
average to carry out a new polarized DIS experiment (see Table I).
This situation was dramatically changed after 1993. Many new
experiments measuring the nucleon and deuteron spin-dependent
structure functions became available. The experimental progress is
certainly quite remarkable in the past years.

Since experimental measurements only cover a limited kinematic
range, an extrapolation to unmeasured $x\to 0$ and $x\to 1$
regions is necessary. At small $x$, a Regge behavior
$g_1(x)\propto x^{\alpha(0)}$ is conventionally assumed in earlier
experimental analyses. The improvement by the new measurements is
two-folds: First, the small $x$ region has been pushed down to the
order $10^{-3}$ in SMC experiments (see Table I). Second, an
extrapolation to the unmeasured small $x$ region is done by
performing a NLO QCD fit rather than by assuming Regge behaviour.
The uncertainties of $\Gamma_1$ which mostly arise from the small
$x$ extrapolation are substantially reduced. From Table I it is
also clear that the EMC experiment has been confirmed by all
successive polarized DIS measurements.

\begin{table}[ht]
\centerline{{\small Table I. Experiments on the polarized
structure functions $g_1^p(x,Q^2),~g_1^n(x,Q^2)$ and
$g_1^d(x,Q^2)$.}} {\footnotesize
\begin{center}
\begin{tabular}{|l c c l l|} \hline
Experiment & Target & $Q^2({\rm GeV}^2)$ range & $x$ range &
$\Gamma^{\rm target}_1(Q^2)$   \\ \hline E80/E130 \cite{E80,E130}
& $p$ & $1<Q^2<10$ & $0.1<x<0.7$ & $\Gamma_1^p(10)=0.17\pm 0.05^*$
\\  E142 \cite{E142}  & $n$ & $1<Q^2<10$ & $0.03<x<0.6$ &
$\Gamma_1^n(2)=-0.031\pm 0.006\pm 0.009$ \\ E143 \cite{E143} &
$p$,$d$ & $1<Q^2<10$ & $0.03<x<0.8$ & $\Gamma_1^p(3)=0.132\pm
0.003\pm 0.009$ \\ & & & & $\Gamma_1^d(3)=0.047\pm 0.003\pm 0.006$
\\ E154 \cite{E154}  & $n$ & $1<Q^2<17$ & $0.014<x<0.7$ &
$\Gamma_1^n(5)=-0.041\pm 0.004\pm 0.006$
\\ E155 \cite{E155} & $p$,$d$ & $1<Q^2<17$ & $0.01<x<0.9$ &
$\Gamma_1^d(5)=0.0266\pm 0.0025\pm 0.0071$
\\ \hline EMC \cite{EMC} & $p$ & $1<Q^2<200$ & $0.01<x<0.7$ &
$\Gamma_1^p(10.7)=0.126\pm 0.010\pm 0.015^{**}$
\\  SMC \cite{SMC98} & $p$,$d$ & $1<Q^2<60$ & $0.003<x<0.7$ &
$\Gamma_1^p(10)=0.120\pm 0.005\pm 0.006\pm 0.014$ \\
 & & & & $\Gamma_1^n(10)=-0.078\pm 0.013\pm 0.008\pm 0.014$ \\ & & & &
$\Gamma_1^d(10)=0.019\pm 0.006\pm 0.003\pm 0.013$
\\ \hline HERMES \cite{HERMES} & $p$,$n$,$d$ & $1<Q^2<10$ & $0.023<x<0.6$ &
$\Gamma_1^n(3)=-0.037\pm 0.013\pm 0.008^\dagger$ \\ \hline
\end{tabular}
\end{center} }
{\footnotesize \noindent $^*$ Obtained by assuming a Regge
behavior $A_1\propto x^{1.14}$ for small $x$.   \\ \noindent
$^{**}$ Combined result of E80, E130 and EMC data. The EMC data
alone give $\Gamma_1^p=0.123\pm 0.013\pm 0.019\,$.
\\ \noindent $^\dagger$
$\int^{0.85}_{0.021}g_1^p(x)dx=0.122\pm 0.003\pm 0.010$ is
obtained by HERMES for the proton target.}
\end{table}
\vskip 0.45cm

Comparing to the original EMC measurement, the statistic and
systematic errors of the combined world average for $\Gamma_1^p$
are substantially reduced. The result is $\Gamma_1^p=(0.12\sim
0.13)\pm 0.007$ at $Q^2=(5\sim 10)\,{\rm GeV}^2$. Consequently,
$\Delta\Sigma= (0.20\sim 0.30)\pm 0.04$. For example,
$\Delta\Sigma=0.25\pm 0.04$ leads to \be \label{newDq} \Delta
u=\,0.81\pm 0.01\,,~~~\Delta d=-0.45\pm 0.01\,,~~~\Delta
s=-0.11\pm 0.01\,.
\en
We will employ (\ref{newDq}) as the benchmarked values for $\Delta
q$ in ensuing discussions.

The Bjorken sum rule evaluated up to $\alpha_s^3$ for three light
flavors is \cite{Lar91}
\be
\Gamma_1^p(Q^2)-\Gamma^n_1(Q^2)=\,{1\over 6}\,{g_A\over
g_V}\left[1- {\alpha_s(Q^2)\over \pi}-{43\over
12}\left({\alpha_s(Q^2)\over\pi}\right)^2-
20.22\left({\alpha_s(Q^2)\over \pi}\right)^3\right].
\en
A serious test of the Bjorken sum rule, which is a rigorous
consequence of QCD, became possible since 1993. The current
experimental results are
\be
{\rm E143}~\cite{E143}: && \Gamma_1^p-\Gamma_1^n=\,0.164\pm
0.021\,, \non \\ {\rm SMC}~\cite{SMC98}: &&
\Gamma_1^p-\Gamma_1^n=\,0.174^{+0.024}_{-0.012}\,,
\en
at $Q^2=5\,{\rm GeV}^2$, to be compared with the prediction
\be
\Gamma_1^p-\Gamma_1^n &=& 0.181\pm 0.003
\en
at the same energies. Therefore, the Bjorken sum rule has been
confirmed by data to an accuracy of 10\% level.

The quark polarization $\Delta q$ is usually determined from the
inclusive data of $g_1$ by assuming SU(3) flavor symmetry.
Moreover, inclusive DIS determines the sum of polarized quark and
antiquark distributions, but not the valence and sea quark spin
distributions. Semi-inclusive polarized experiments in principle
allow the determination of $\Delta q$ for each flavor and
disentangle valence and sea polarizations separately \cite{Frank}.
Hence, SU(3) flavor symmetry can be tested by comparing the
measured first moments of the flavor distributions to the SU(3)
predictions \cite{HERMESsemi}. Semi-inclusive data are available
from SMC \cite{SMCsemi} and HERMES \cite{HERMESsemi} (see Table
II). In order to present the experimental results, we digress for
a moment to adopt a different definition for quark spin densities
here: $\Delta q_s(x)=q_s^\up(x)-q _s^\down(x)$ and
$\Delta\bar{q}(x)=\bar{q}^\up(x)-\bar{q}^\down(x)$, where $\Delta
q_s(x)=\Delta q(x)-\Delta q_v(x)$ is the sea spin distribution for
the quark flavor $q$. The SMC analysis is based on the assumption
of SU(3) flavor symmetric sea: $\Delta \bar u(x)=\Delta\bar
d(x)=\Delta\bar s(x)=\Delta u_s(x)=\Delta d_s(x)=\Delta s(x)$,
while the HERMES results shown in Table II rely on the assumption
of flavor independent polarization:
\be
{\Delta u_s(x)\over u_s(x)}={\Delta d_s(x)\over d_s(x)}={\Delta
s(x)\over s(x)}={\Delta\bar u(x)\over \bar u(x)}={\Delta\bar
d(x)\over \bar d(x)}={\Delta\bar s(x)\over \bar s(x)}.
\en
Note that the ansatz\footnote{This ansatz is generally not
fulfilled by the assumption of flavor independent polarization
made by HERMES \cite{HERMESsemi}.} of $\Delta \bar q(x)=\Delta
q_s(x)$ has to be made in both experiments in order to extract the
valence quark polarization $\Delta q_v$ from the measurement of
$\Delta q$ and $\Delta\bar q$, i.e., $\Delta q_v(x)=\Delta
q(x)-\Delta\bar q(x)$. However, one caveat has to be mentioned:
{\it A priori} the antiqaurk spin $\Delta\bar{q}$ can be different
from the sea polarization $\Delta q_s$ if they are not produced
from gluons. For example, antiquarks are not polarized in the
model of \cite{ChengLi96}. Under the assumption of SU(3) symmetric
sea polarization, we are led to the predictions $g_A^3=\Delta
u_v-\Delta d_v$, $g_A^8=\Delta u_v+\Delta d_v$ and hence $\Delta
u_v=2F=0.93\pm 0.02$ and $\Delta d_v=F-D=-0.34\pm 0.02$. Note that
the valence polarization $\Delta q_v$ should be scale independent.

\begin{table}[ht]
{\small Table II. The SMC \cite{SMCsemi} and HERMES
\cite{HERMESsemi} results for the first moments of valence and sea
spin distributions.} {
\begin{center}
\begin{tabular}{| l| c c |} \hline
 & SMC ($Q^2=10\,{\rm GeV}^2$)  & HERMES ($Q^2=2.5\,{\rm GeV}^2$)
  \\ \hline
$\Delta u_v$ & $0.77\pm 0.10\pm 0.08$ & $0.57\pm 0.05\pm 0.08$ \\
$\Delta d_v$ & $-0.52\pm 0.14\pm 0.09$ & $-0.22\pm 0.11\pm 0.13$
\\ $\Delta\bar u$ & $0.01\pm 0.04\pm 0.03$ & $-0.01\pm 0.02\pm
0.03$ \\ $\Delta \bar d$ & $0.01\pm 0.04\pm 0.03$ & $-0.01\pm
0.03\pm 0.04$ \\ $x\Delta u_v$ & $0.155\pm 0.017\pm 0.010$ &
$0.13\pm 0.01\pm 0.01$ \\ $x\Delta d_v$ & $-0.056\pm 0.026\pm
0.011$ & $-0.02\pm 0.02\pm 0.02$ \\ \hline
\end{tabular}
\end{center} }
\end{table}
\vskip 0.45cm

The measurement of the gluon spin $\Delta G$ by all possible means
is very important both theoretically and experimentally (see
\cite{Cheng96b} for various processes sensitive to the gluon spin
distributions). A global fit to the present inclusive DIS data of
$g_1(x)$ cannot even fix the sign of $\Delta G$ decisively (see
Sec. 3.3), not mentioning its magnitude. One way of measuring
$\Delta G(x)$ directly is via the photon gluon fusion process
occurred in the semi-inclusive DIS reaction. A recent HERMES
\cite{HERMESdG} measurement of the longitudinal spin asymmetry in
photoproduction of pairs of hadrons with high transverse momentum
indicates that $\la \Delta G(x)/G(x)\ra=0.41\pm 0.18\pm 0.03$ at
$\la x\ra=0.17$ to LO QCD. Hence $\Delta G(x)$ is found to be
positive in the intermediate $x$ region.

\section{Theoretical Progress}
\setcounter{equation}{0}
In my opinion there are four important
progresses in theory:
\begin{itemize}
\item The role played by the gluon to the first moment of $g_1$ is
clarified. There are two extreme factorization schemes of
interest.
\item First-principles calculations of the quark spin and orbital angular momentum
by lattice QCD became available.
\item A complete and consistent NLO analysis of $g_1$ data became
possible.
\item Evoluation and gauge dependence of the quark orbital angular momentum
are explored.

\end{itemize}
\subsection{Anomalous gluon and sea quark interpretations}
\subsubsection{Anomalous gluon interpretation}

We see from Sec. II that the polarized DIS data indicate that the
fraction of the proton spin carried by the light quarks inside the
proton is $\Delta\Sigma= (0.20\sim 0.30)$ and the strange-quark
polarization is $\Delta s\approx -0.10$ at $Q^2=(5\sim 10)\,{\rm
GeV}^2$. The question is what kind of mechanism can generate a
sizeable and negative sea polarization. It is difficult, if not
impossible, to accommodate a large $\Delta s$ in the naive parton
model because massless quarks and antiquarks created from gluons
have opposite helicities owing to helicity conservation. This
implies that sea polarization for massless quarks cannot be
induced perturbatively from hard gluons, irrespective of gluon
polarization.  It is also unlikely that the observed $\Delta s$
comes solely from nonperturbative effects or from chiral-symmetry
breaking due to nonvanishing quark masses.

As an attempt to understand the polarized DIS data, we consider
QCD corrections to the polarized proton structure function
$g_1^p(x)$. To the next-to-leading order (NLO) of $\alpha_s$, the
expression for $g_1^p(x)$ is
\be
g_1^p(x,Q^2) &=& {1\over 2}\sum e_q^2\Big[ \Delta
C_q(x,\alpha_s)\otimes \Delta q(x,Q^2)+\Delta
C_G(x,\alpha_s)\otimes\Delta G(x,Q^2)\Big]\non\\ &=& {1\over
2}\sum e_q^2\Big[ \Delta q^{(0)}(x,Q^2)+\Delta
q^{(1)}(x,Q^2)+\Delta q_s^G(x,Q^2)   \\ && +\Delta
C_q^{(1)}(x,\alpha_s)\otimes\Delta q^{(0)}(x,Q^2)+ \Delta
C_G^{(1)}(x,\alpha_s)\otimes\Delta G(x,Q^2)+\cdots\Big], \non
\en
where uses of $\Delta C_q^{(0)}(x)=\delta(1-x)$ and $\Delta
C_G^{(0)}(x)=0$ have been made, $\otimes$ denotes the convolution
\be
f(x)\otimes g(x)=\int^1_x {dy\over y}\,f\left({x\over y}\right)
g(y),
\en
and $\Delta C_q$, $\Delta C_G$ are short-distance quark and gluon
coefficient functions, respectively. More specifically, $\Delta
C^{(1)}_G$ arises from the hard part of the polarized photon-gluon
cross section, while $\Delta C^{(1)}_q$ from the short-distance
part of the photon-quark cross section. Contrary to the
coefficient functions, $\Delta q_s^G(x)$ and $\Delta q^{(1)}(x)$
come from the soft part of polarized photon-gluon and photon-quark
scatterings, respectively. Explicitly, they are given by
\be
\label{dqG}  \Delta q^{(1)}(x,Q^2)=\Delta
\phi^{(1)}_{q/q}(x)\otimes \Delta q^{(0)}(x,Q^2), \qquad \Delta
q_s^G(x,Q^2)=\Delta\phi_{q/G}^{(1)}(x)\otimes\Delta G(x,Q^2),
\en
where $\Delta \phi_{j/i}(x)$ is the polarized distribution of
parton $j$ in parton $i$. Diagrammatically,
$\Delta\phi_{q/q}^{(1)}$ and $\Delta\phi_{q/G}^{(1)}$ are depicted
in Fig.~1.

\begin{figure}[ht]
\vspace{-3cm} \hskip 1cm
  \psfig{figure=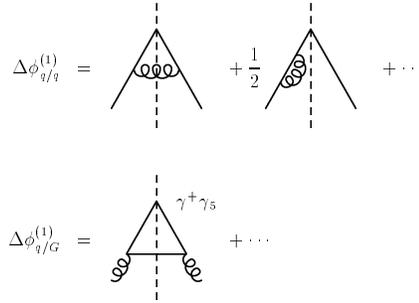,width=13cm}
\vspace{-11cm}
    \caption[]{\small Diagrams for the quark spin distributions inside the
parton: $\Delta\phi^{(1)}_{q/q}$ and $\Delta\phi^{(1)}_{q/G}$.}
    \label{fig1}
\end{figure}

The photon-gluon scattering box diagram is ultraviolet finite but
it depends on the choice of infrared and collinear regulators.
However, the hard part of the box diagram should be soft-regulator
independent. In the improved parton model, this is done by
introducing a factorization scale $\mu_{\rm fact}$ so that the
region $k^2_\perp\gsim \u^2$ contributes to the hard photon-gluon
cross section. The results are (see e.g. \cite{Cheng96b})
\be
\label{DGCI} \Delta C_G^{(1)}(x,Q^2,\u^2)_{\rm CI} &=&
(2x-1)\left(\ln{Q^2\over \u^2} +\ln{1-x\over x}-1\right), \non \\
\int^1_0 dx\Delta C^{(1)}_G(x)_{\rm CI}&=& -{\alpha_s\over 2\pi},
\en
where the reason for introducing the subscript ``CI" will become
clear below. Hence,
\be
\label{Gamma1p} \Gamma_1^p(Q^2)\equiv\int^1_0
dxg_1^p(x,Q^2)=\,{1\over 2}\left(1-{ \alpha_s\over
\pi}\right)\sum_i\left[\Delta q_i(Q^2)_{\rm
CI}-{\alpha_s(Q^2)\over 2\pi}\Delta G(Q^2)\right].
\en
The $(1-{\alpha_s\over \pi})$ term in  Eq. (\ref{Gamma1p}) comes
from the QCD loop correction, while the $\alpha_s\Delta G$ term
arises from the box diagram of photon-gluon scattering. If the
gluon polarization inside the proton is positive, a partial
cancellation between $\Delta q_{\rm CI}$ and ${\alpha_s\over
2\pi}\Delta G$ will explain why the observed $\Gamma_1^p$ is
smaller than what naively expected from the Ellis-Jaffe sum rule.
Note that unlike the usual QCD corrections, the QCD effect due to
photon-gluon scattering is very special: The term $\alpha_s\Delta
G$ is conserved to the leading-order QCD evolution; that is,
$\Delta G$ grows with $\ln Q^2$, whereas $\alpha_s$ is inversely
proportional to $\ln Q^2$.

As a consequence, Eq. (\ref{newDq}) is modified to
\be
\label{deltaq'} \Delta u_{\rm CI}-{\alpha_s\over 2\pi}\Delta G &=&
0.81\pm 0.01\,, \non
\\ \Delta d_{\rm CI}-{\alpha_s\over 2\pi}\Delta G &=& -0.45\pm
0.01\,,
\\ \Delta s_{\rm CI}-{\alpha_s\over 2\pi}\Delta G &=& -0.11\pm
0.01\,,  \non
\en
and
\be \label{Sigma'}
g_A^0=(\Delta u+\Delta d+\Delta s)_{\rm
CI}-{3\alpha_s\over 2\pi}\Delta G=\,0.25 \pm 0.04
\en
at $Q^2=5\sim 10\,{\rm GeV}^2$. Eqs. (\ref{deltaq'}) and
(\ref{Sigma'}) imply that in the presence of anomalous gluon
contributions, $\Delta\Sigma_{\rm CI}$ is not necessarily small
while $\Delta s_{\rm CI}$ is not necessarily large. In the absence
of sea polarization and in the framework of perturbative QCD, it
is easily seen that $\Delta G\sim \Delta s(2\pi/\alpha_s)\sim 2.5$
at $Q^2= 10\,{\rm GeV}^2$ and $\Delta\Sigma_{\rm CI}\sim 0.58\,.$
It thus provides a nice and simple solution to the proton spin
puzzle: This improved parton picture is reconciled, to a large
degree, with the constituent quark model and yet explains the
suppression of $\Gamma_1^p$, provided that $\Delta G$ is positive
and large enough. This anomalous gluon interpretation of the
observed $\Gamma_1^p$, as first proposed in \cite{Efr,AR,CCM} (see
also \cite{Leader88}), looks appealing and has become a popular
explanation since 1988.

Some historical remarks are in order:
\begin{itemize}
\item Long before the EMC experiment, there already existed three theoretical
calculations of the photon-gluon box diagram. Kodaira
\cite{Kodaira} was the first one to compute the moments of the
structure functions $g_{1,2}$. Since he worked in the OPE
framework, there is no decomposition of $g_A^0$ in terms of quark
and gluon spin components. The anomalous gluonic contribution to
$\Gamma_1^p$ was first put forward by Lam and Li \cite{Lam} in
1982. A calculation of the gluonic coefficient function using the
dimensional regularization was first made by Ratcliffe \cite{Rat}.
\item The original results for the photon-gluon scattering cross
section obtained by \cite{Efr,AR,CCM} are not perturbative QCD
reliable as they depend on the choice of soft regulators. The
first moment of $\Delta C_G^{(1)}(x)$ is equal to
$-\alpha_s/(2\pi)$ in \cite{CCM} but vanishes in the work of
\cite{AR,Rat}. After the soft part below the factorization scale
$\u$ is removed, the gluon coefficient function is given by Eq.
(\ref{DGCI}) which is soft-cutoff independent.

\end{itemize}

\subsubsection{Sea quark interpretation}

According to the operator product expansion (OPE), the moments of
structure functions can be expressed in terms of hard coefficients
which are calculable by perturbative QCD and forward matrix
elements of local gauge-invariant operators which are
nonperturbative in nature:
\be
\int_0^1dx\,x^{n-1}g(x)=\sum _iC^n_i(Q^2)\la N|O_i^n(0)|N\ra. \en
It turns out that there is no gauge-invariant twist-2, spin-1
local gluonic operator for the first moment of $g_1(x)$ (see e.g.
\cite{JM}). Here we face a dilemma here: On the one hand, the
anomalous gluon interpretation sounds more attractive and is able
to reconcile to a large degree with the conventional quark model;
on the other hand, the sea-quark interpretation of $\Gamma_1^p$
relies on a more solid theory of the OPE. In fact, these two
popular explanations for the $g_1^p$ data have been under hot
debate over many years before 1995. Though the OPE approach is
model independent, it faces the questions of what is the deep
reason for the absence of gluonic contributions to $\Gamma_1^p$
and how are we going to understand a large and negative
strange-quark polarization ?

\subsubsection{Factorization scheme dependence}

In spite of much controversy on the aforementioned issue, this
dispute was actually resolved almost a decade ago \cite
{BQ,Manohar}. The key point is that a different interpretation for
$\Gamma^p_1$ corresponds to a different factorization definition
for the quark spin density and the hard photon-gluon cross
section. The choice of the ``ultraviolet" cutoff for soft
contributions specifies the factorization convention.\footnote{It
is misleading to identify the regularization scheme for soft
divergences, e.g. the off-shell scheme, with the factorization
scheme; the former is merely employed to get rid of infrared and
collinear divergences appearing in the calculation of partonic
cross sections. However, the hard gluon coefficient functions are
soft-cutoff independent, and they depend on the ultraviolet
regulator on the triangle diagram for defining the polarized quark
distribution inside the gluon. In other words, it is the choice of
ultraviolet regulator rather than the soft cutoff that specifies
the factorization scheme.} More specifically, since
$\Delta\phi^{(1)}$ in Eq. (3.1) is ultraviolet divergent, it is
clear that, just like the case of unpolarized deep inelastic
scattering, the coefficient functions $\Delta C_q$ and $\Delta
C_G$ depend on how the parton spin distributions $\Delta
\phi^{(1)}_{j/i}$ are defined, or how the ultraviolet regulator is
specified on $\Delta\phi^{(1)}$. That is, the ambiguities in
defining $\Delta\phi^{(1)}_{q/q}$ and $\Delta\phi^{(1)}_{q/G}$ are
reflected on the ambiguities in extracting $\Delta C^{(1)}_q$ and
$\Delta C^{(1)}_G$. Consequently, the decomposition of the
photon-gluon and photon-quark cross sections into the hard and
soft parts depends on the choice of the factorization scheme and
the factorization scale $\mu$. Of course, the physical quantity
$g^p_1(x)$ is independent of the factorization prescription (for a
review on the issue of factorization, see \cite{Cheng96b}).

However, the situation for the polarized DIS case is more
complicated: In addition to all the ambiguities that spin-averaged
parton distributions have, the parton spin densities are subject
to two extra ambiguities, namely, the axial anomaly and the
definition of $\gamma_5$ in $n$ dimension. It is well known that
the polarized triangle diagram for $\Delta\phi^{(1)}_{q/G}$ (see
Fig.~1) has an axial anomaly. There are two extreme ultraviolet
regulators of interest. One of them, which we refer to as the
chiral-invariant (CI) factorization scheme (sometimes called as
the ``jet scheme" \cite{Leader98pl}, the ``parton model scheme"
\cite{Bass99} or the ``$k_\perp$ cut-off scheme"), respects chiral
symmetry and gauge invariance but not the axial anomaly. This
corresponds to a direct brute-force cutoff $\sim \mu$ on the
$k_\perp$ integration in the triangle diagram ( i.e.
$k^2_\perp\lsim \mu^2$) with $k_\perp$ being the quark transverse
momentum perpendicular to the virtual photon direction. Since the
gluonic anomaly is manifested at $k_\perp^2\to\infty$, it is
evident that the spin-dependent quark distribution [i.e. $\Delta
q^{(1)}(x)$] in the CI factorization scheme is anomaly-free. Note
that this is the $k_\perp$-factorization scheme employed in the
usual improved parton model \cite{Efr,AR,CCM}.

   The other ultraviolet cutoff on the triangle diagram of Fig.~1, as
employed in the approach of the OPE, is chosen to satisfy gauge
symmetry and the gluonic anomaly. As a result, chiral symmetry is
broken in this gauge-invariant (GI) factorization scheme and the
sea polarization is perturbatively induced from hard gluons via
the anomaly. This perturbative mechanism for sea quark
polarization is independent of the light quark masses. A
straightforward calculation gives \cite{Bass,Cheng96a}
\be
\Delta\phi_{q/G}^{(1)}(x)_{\rm GI}=\Delta\phi_{q/G}^{(1)}(x)_{\rm
CI} -{\alpha_s\over \pi}(1-x),   \label{phi}
\en
where the term ${\alpha_s\over\pi}(1-x)$ originates from the QCD
anomaly arising from the region where $k_\perp^2\to\infty$. Two
remarks are in order. First, this term has been erroneously
claimed\footnote{Some misconceptions in the literature about the
$\MS$ scheme have to be clarified. It has been argued \cite{Ball}
that the GI scheme is pathologic and inappropriate since $\Delta
C^{(1)}_G(x)_{\rm GI}$, which is ``hard" by definition, contains
an unwanted ``soft" term proportional to $(1-x)$ [see Eq.
(\ref{deltag})]. The cross section of the photon-gluon box diagram
contains a term proportional to $(1-x)$ if the infrared and
collinear divergences are regulated by the quark mass or by the
dimensional regulator. If the ultraviolet regulator for the
triangle diagram is chirality-preserving, the $(1-x)$ term, which
arises from the region where $k_T^2\sim m_q^2$, does not
contribute to the hard gluon coefficient, as it should be.
However, if the ultraviolet regulator preserves the axial anomaly
and gauge invariance, for example, the $\MS$ regulator, chirality
will be broken and the axial anomaly is absorbed in the quark spin
density. It turns out that the effect of the axial anomaly, which
is manifested at $k_T^2\sim \mu_{\rm fact}^2$ (the factorization
scale), has the $x$ dependence of the $(1-x)$ form. This explains
why $\Delta C_G$ in the $\MS$ scheme has a term proportional to
$(1-x)$ because the axial-anomaly effect must be subtracted from
the previous gluon coefficient function computed in the
chiral-invariant scheme. As pointed out in
\cite{Cheng96a,Cheng96b} and again in \cite{Cheng98}, the $(1-x)$
term in the gluonic coefficient function in the $\MS$ scheme is
purely ``hard", contrary to what has been claimed previously.} in
some literature \cite{Mank,Ball,Lampe,Leader98pl} to be a soft
term coming from $k_\perp^2\sim m_q^2$. Second, although the quark
spin distribution inside the gluon $\Delta\phi^{(1)}_{q/G}(x)$
cannot be reliably calculated by perturbative QCD, its difference
in GI and CI schemes is trustworthy in QCD. Since the polarized
valence quark distributions are $k_\perp$-factorization scheme
independent, the total quark spin distributions in GI and CI
schemes are related via Eqs. (\ref{dqG}) and (\ref{phi}) to be
\cite{Cheng97}
\be
\Delta q(x,Q^2)_{\rm GI}=\Delta q(x,Q^2)_{\rm
CI}-{\alpha_s(Q^2)\over \pi}\,(1-x)\otimes\Delta G(x,Q^2).
\label{deltaq}
\en
For a derivation of this important result based on a different
approach, namely, the nonlocal light-ray operator technique, see
M\"uller and Teryaev \cite{Muller}.

The axial anomaly in the box diagram for polarized photon-gluon
scattering also occurs at $k_\perp^2\to\infty$, more precisely, at
$k^2_\perp=[(1-x)/4x]Q^2$ with $x\to 0$. It is natural to expect
that the axial anomaly resides in the gluon coefficient function
$\Delta C^{(1)}_G$ in the CI scheme, whereas its effect in the GI
scheme is shifted to the quark spin density. Since $\Delta
C^{(1)}_G(x)$ is the hard part of the polarized photon-gluon cross
section, which is sometimes denoted by $g_1^G(x)$, the polarized
structure function of the gluon target, we have
\be
\Delta C_G^{(1)}(x)=\,g_1^G(x)-\Delta\phi_{q/G}^{(1)}(x).
\en
It follows from Eqs.~(\ref{phi}) and (\ref{deltaq}) that
\be
\Delta C^{(1)}_G(x)_{\rm GI}=\Delta C^{(1)}_G(x)_{\rm
CI}+{\alpha_s\over \pi} (1-x).   \label{deltag}
\en

The first moments of $\Delta C_G(x)$, $\sum_q\Delta q(x)$ and
$g_1^p(x)$ are given by
\be
&& \int^1_0 dx\Delta C^{(1)}_G(x)_{\rm GI}=0, \qquad \int^1_0
dx\Delta C^{(1)}_G(x)_{\rm CI}=-{\alpha_s\over 2\pi}, \non   \\ &&
\Delta\Sigma_{\rm GI}(Q^2)=\,\Delta\Sigma_{\rm
CI}(Q^2)-{n_f\alpha_s(Q^2) \over 2\pi}\Delta G(Q^2),
\en
and
\be \label{Deltap}
\Gamma^p_1 &=& {1\over 2}\sum
e_q^2\left(\Delta q_{\rm CI}(Q^2)-{\alpha_s(Q^2)\over 2\pi}\Delta
G(Q^2)\right) \non
\\ &=& {1\over 2}\sum e_q^2\Delta q_{\rm GI}(Q^2),
\en
where we have neglected contributions to $g_1^p$ from
$\Delta\phi^{(1)}_{q/q}$ and $\Delta C_q^{(1)}$. Note that
$\Delta\Sigma_{\rm GI}(Q^2)$ is equivalent to the singlet axial
charge $\la p,s|J_\mu^5 |p,s\ra$. The well-known results
(\ref{deltaq}-\ref{Deltap}) indicate that $\Gamma_1^p$ receives
anomalous gluon contributions in the CI factorization scheme (e.g.
the improved parton model), whereas hard gluons do not play any
role in $\Gamma_1^p$ in the GI scheme such as the OPE approach.
From (\ref{Deltap}) it is evident that the sea quark or anomalous
gluon interpretation for the suppression of $\Gamma_1^p$ observed
experimentally is simply a matter of convention \cite{BQ}.

The $\overline{\rm MS}$ scheme is the most common one chosen in
the GI factorization convention. The so-called Adler-Bardeen (AB)
factorization scheme often adopted in the literature is obtained
from the GI scheme by adding the $x$-independent term
$-\alpha_s/(2\pi)$ to $\Delta C_G^{\rm GI}$ via
\be
\Delta C_G(x)_{\rm AB}=\Delta C_G(x)_{\rm GI}-{\alpha_s\over
2\pi}, \en while
\be
\Delta q(x,Q^2)_{\rm AB}=\Delta q(x,Q^2)_{\rm
GI}+{\alpha_s(Q^2)\over 2\pi}\int^1_x{dy\over y}\Delta G(y,Q^2).
\en
In general, one can define a family of schemes labelled by the
parameter $a$ \cite{Zijl,Leader98pl}:
\be
\Delta q(x)_a=\Delta q_{\rm GI}(x)+{\alpha_s\over
2\pi}\left[(2x-1)(a-1)+2(1-x)\right]\otimes \Delta G(x),
\en
which satisfy the relation
\be
\Delta\Sigma_a=\Delta\Sigma_{\rm GI}+{3\alpha_s\over 2\pi}\Delta
G,
\en
to the first moment, but differ in their higher moments. The AB
scheme corresponds to $a=2$ and the CI scheme to $a=1$. Since the
$x$ dependence of the axial-anomaly contribution in the quark
sector is fixed to be $(1-x)$, it is obvious that all the schemes
in this family including the AB scheme cannot consistently put all
hard anomaly effects into gluonic coefficient functions unless
$a=1$, contrary to the original claim made in \cite{Ball}.

Finally, it should be stressed that the quark coefficient function
$\Delta C_q^{(1)}(x)$ in the dimensional regularization scheme is
subject to another ambiguity, namely, the definition of $\gamma_5$
in $n$ dimension used to specify the ultraviolet cutoff on
$\Delta\phi_{q/q}^{(1)}$ (see Fig.~1). For example, $\Delta
C_q^{(1)}(x)$ calculated in the $\gamma_5$ prescription of 't
Hooft and Veltman, Breitenlohner and Maison (HVBM) is different
from that computed in the dimension reduction scheme. The result
\be
\Delta C_q^{(1)}(x)=C_q^{(1)}(x)-{2\alpha_s\over 3}(1+x)
\en
usually seen in the literature is obtained in the HVBM scheme,
where $C_q(x)$ is the unpolarized quark coefficient function. Of
course, the quantity $\Delta q^{(1)}(x)+\Delta C_q(x)\otimes\Delta
q^{(0)}(x)$ and hence $g_1^p(x)$ is independent of the definition
of $\gamma_5$ in dimensional regularization.

\subsubsection{A brief summary}
In addition to all the ambiguities that spin-averaged parton
distributions have, the parton spin densities are subject to two
extra ambiguities, namely, the axial anomaly and the definition of
$\gamma_5$ in $n$ dimension. There are two extreme cases of
interest: Either the hard axial anomaly is manifested in the
matrix elements of the quark current (GI scheme) or it is absorbed
in the gluonic coefficient function so that the quark matrix
element is anomaly-free (CI scheme). It should be stressed that in
the so-called AB scheme, not all hard anomaly effects are lumped
into $\Delta C_G$ and hence the corresponding quark matrix element
is not anomaly-free. Of course, it appears that the CI scheme is
close to the intuitive parton picture as the quark spin
distribution which does not contain hard contributions from the
anomaly is scale independent. (The price to be paid is that
$\Delta q_{\rm CI}$ cannot be expressed as the matrix element of a
local gauge-invariant operator and hence it is difficult to
compute by lattice QCD. Also, {\it a priori} there is no reason
that $\Delta\Sigma$ should have a simple quark interpretation
\cite{Smith}.) Nevertheless, physically and mathematically GI and
CI schemes are equivalent.

Two remarks are in order. (i) It is worth emphasizing that
although the suppression of $\Gamma_1^p$ can be accommodated by
anomalous gluon and/or sea quark contributions, no quantitative
prediction of $\Gamma_1^p$ can be made. An attempt of explaining
the smallness of $g_A^0$ has been made in the large-$N_c$ approach
\cite{Brod88}. (ii)  So far we have focused on the perturbative
part of the axial anomaly. The perturbative QCD result
(\ref{deltaq}) indicates that the difference $\Delta q_s^{\rm
GI}-\Delta q_s^{\rm CI}$ ($\Delta q=\Delta q_v+\Delta q_s$ with
the valence polaization $\Delta q_v$ being scheme independent) is
induced perturbatively from hard gluons via the anomaly mechanism
and its sign is predicted to be negative. By contrast, $\Delta
q_s^{\rm CI}(x)$ can be regarded as an intrinsic sea-quark spin
density produced nonperturbatively. The well-known solution to the
$U_A(1)$ problem in QCD involves two important ingredients: the
QCD anomaly and the QCD vacuum with a nontrivial topological
structure, namely the $\theta$-vacuum constructed from instantons
which are nonperturbative gluon configurations. Since the
instanton-induced interactions can flip quark helicity, in analog
to the baryon-number nonconservation induced by the 't Hooft
mechanism, the quark-antiquark pair created from the QCD vacuum
via instantons can have a net helicity. It has been suggested that
this mechanism of quark helicity nonconservation provides a
natural and nonperturbative way of generating negative sea-quark
polarization \cite{Dor}.

In retrospect, the dispute among the anomalous gluon and sea-quark
explanations of the suppression of $\Gamma_1$ mostly before 1996
is considerably unfortunate and annoying since the fact that
$g_1(x)$ is independent of the definition of the quark spin
density and hence the choice of the factorization scheme due to
the axial-anomaly ambiguity is presumably well known to all the
practitioners in the field, especially to those QCD experts
working in the area.

\subsection{Lattice calculation of the proton spin content}
The present lattice calculation is starting to shed light on the
proton spin contents. An evaluation of $\Delta q_{\rm GI}$, the
gauge-invariant quark spin component defined by $\Delta q_{\rm
GI}=\la p,s|\bar{q}\gamma^\mu\gamma_5 q|p,s\ra s_\mu$, involves a
disconnected insertion in addition to the connected insertion (see
Fig.~2; the infinitely many possible gluon lines and additional
quark loops are implicit). The sea-quark spin contribution comes
from the disconnected insertion.

There are four lattice calculations done by Dong et al.
\cite{Dong}, Fukugita et al. \cite{Fukugita}, G\"ockeler et al.
\cite{Gockeler1,Gockeler2} in the quenched approximation and
G\"usken et al. \cite{Gusken} in full lattice QCD. Note that the
disconnected contribution is not evaluated in
\cite{Gockeler1,Gockeler2}. From Table III it is clear that the
lattice results for $g_A^0$ and $\Delta s$ are in agreement with
experiment, while the full lattice QCD calculations for
$g_A^3,~\Delta u$ and $\Delta d$ are too small compared to
experiment. In particular, there is a 30\% discrepancy between the
lattice QCD estimate of $g_A^3$ \cite{Gusken} and the experimental
value. This points to the presence of sizeable higher order or
even nonperturbative contributions to the renormalized factor
$Z_A^{\rm NS}$ on the non-singlet current.

\begin{figure}[ht]
\vspace{+1cm}
    \centerline{\psfig{figure=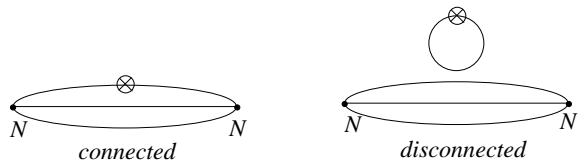,width=7.5cm}}
    \caption[]{\small Connected and disconnected insertions arising from the
    flavor-singlet axial-vector current.}
    \label{fig4}

\end{figure}

\begin{table}[ht]
{\small Table III. Quark spin contents of the proton from lattice
calculations. Experimental results are taken from (\ref{newDq})
for the first moments of quark spin distributions and Table II for
the second moments$^\dagger$ $\Delta^{(1)}q_v$ ($=\int^1_0 x\Delta
q_v(x)dx$) of valence quark spin densities.}
\begin{center}
\begin{tabular}{|c|l l l l | l|} \hline
 & Dong et al. & Fukugita et al. & G\"ockeler et al.$^*$ & G\"usken et al.  &
 Expt  \\  \hline
$g_A^0$ & 0.25(12) & 0.18(10) & & 0.20(12) & 0.25(4)  \\ $g_A^3$ &
1.20(10) & 0.985(25) & 1.24(10) & 0.907(20) & 1.2670(35)  \\
$g_A^8$ & 0.61(13) & & & 0.484(18) & 0.585(25)
\\ $\Delta u$ & 0.79(11) & 0.638(54) & 0.84(5) & 0.62(7) & 0.81(1)
\\ $\Delta d$ &
-0.42(11) & -0.347(46) & -0.24(2) & -0.29(6) & -0.45(1) \\ $\Delta
s$ & -0.12(1) & -0.109(30) & & -0.12(7) & -0.11(1)\\ $\Delta^{(1)}
u_v$ & & & 0.198(8) & & 0.169(22) \cite{SMCsemi} \\  & & & & &
0.13(1) \cite{HERMESsemi} \\ $\Delta^{(1)} d_v$ & & & -0.0477(33)
& & -0.055(29) \cite{SMCsemi} \\ & & & & & -0.02(3)
\cite{HERMESsemi}
 \\ \hline
\end{tabular}
\end{center}
{\footnotesize  \noindent $^*$Since the disconnected contributions
are not calculated in \cite{Gockeler1,Gockeler2}, $\Delta q$
receives contributions only from the valence component. The
lattice results for $\Delta^{(1)}q_v$, which are presumably scale
independent, are obtained in \cite{Gockeler2} at the scale
$\mu^2=4\,{\rm GeV}^2$.}
\end{table}

As for the chiral-invariant quantity $\Delta\Sigma_{\rm CI}$, it
involves the matrix element of $\tilde{J}^+_5$ in light-front
gauge where $\tilde J_{\mu 5}$ is an anomaly-free singlet axial
vector current and hence sizeable gauge configurations are needed
in lattice calculations for $\Delta\Sigma_{\rm CI}$. Nevertheless,
it is conceivable to have lattice results for $\Delta G$ and
$\Delta q_{\rm CI}$ soon in the near future. The lattice
calculation of the quark total angular momentum was also available
recently, see Sec. 3.4.

\subsection{NLO analysis of polarization data}
The experimental data of $g_1(x,Q^2)$ taken at different $x$-bin
correspond to different ranges of $Q^2$; that is, $Q^2$ of the
data is $x$-bin dependent. Hence, it is desirable to evolve the
data to a common value of $Q^2$ in order to determine the moments
of $g_1$ and test the Bjorken sum rule. Because of the
availability of the two-loop polarized splitting functions $\Delta
P^{(1)}_{ij}(x)$ \cite{Zijl,Mert}, it became possible to embark on
a full next-to-leading order (NLO) analysis of the experimental
data of polarized structure functions by taking into account the
measured $x$ dependence of $Q^2$ at each $x$ bin. The NLO analyses
have been performed in the $\overline{\rm MS}$ scheme (a family of
the GI scheme)
\cite{Gluck96,Gehr96,SMC97,E154,Leader98pr,Florian98,Gordon,SMC98,Bourrely98,Leader99,Tatur,Goto,Ghosh},
the Adler-Bardeen (AB) scheme
\cite{Ball,Altarelli97,E154,Gordon,Leader99,SMC98} and the CI
scheme \cite{Leader98pl,Leader99}. Of course, physical quantities
such as the polarized structure function $g_1(x)$ are independent
of choice of the factorization convention. Physically, the
spin-dependent valence quark and gluon distributions should be the
same in all factorization schemes.

The $Q^2$ dependence of parton spin densities is determined by the
spin-dependent DGLAP equations:
\be
&& {d\over dt}\Delta q_{\rm NS}(x,t)=\,{\alpha_s(t)\over
2\pi}\Delta P_{qq}^ {\rm NS}(x)\otimes\Delta q_{\rm NS}(x,t),
\non \\ && {d\over dt}\left(\matrix{\Delta q_{\rm S}(x,t)   \cr
\Delta G(x,t) \cr} \right)=\,{\alpha_s(t)\over
2\pi}\left(\matrix{\Delta P_{qq}^{\rm S}(x) & 2n_f\Delta P_{qG}(x)
\cr  \Delta P_{Gq}(x) & \Delta P_{GG}(x) \cr}
\right)\otimes\left(\matrix {\Delta q_{\rm S}(x,t)   \cr   \Delta
G(x,t) \cr} \right),
\en
with $t=\ln(Q^2/\Lambda^2_{_{\rm QCD}})$,
\be
\Delta q_{\rm NS}(x)=\Delta q_i(x)-\Delta q_j(x),~~~~~\Delta
q_{\rm S}(x)= \sum_i\Delta q_i(x),
\en
and
\be
\Delta P_{ij}(x)=\Delta P_{ij}^{(0)}(x)+{\alpha_s\over 2\pi}\Delta
P_{ij}^{ (1)}(x)+\cdots.
\en
The spin-dependent anomalous dimensions are defined as
\be
\Delta\gamma_{ij}^n=\int^1_0\Delta
P_{ij}(x)x^{n-1}dx=\Delta\gamma_{ij}^{(0), n}+{\alpha_s\over
2\pi}\Delta\gamma_{ij}^{(1),n}+\cdots.
\en
To the NLO, $\Delta P_{qq}^{(1)}$ and $\Delta P_{qG}^{(1)}$ were
calculated in the $\overline{\rm MS}$ scheme by Zijlstra and van
Neerven \cite{Zijl}. However, the other two polarized splitting
functions $\Delta P_{Gq}^{(1)}$ and $\Delta P_{GG}^{(1)}$ were not
available until 1995 \cite{Mert}. The recent analysis \cite{Goto}
shows that the NLO $\chi^2$ is significantly smaller than that of
LO, indicating the necessity of NLO fit of data in practice.

Since the unpolarized parton densities are mostly parameterized in
the $\MS$ scheme and the two-loop splitting functions are
available in the same scheme, it is quite natural to perform the
NLO analysis in the $\MS$ scheme. In principle one can also work
in any other factorization scheme. The splitting functions $\Delta
P_{ij}^{(1)}$ in the CI scheme, for example, is obtained by
applying Eq. (\ref{deltaq}) to the spin-dependent DGLAP equation,
see \cite{Muller,Cheng98}. It is worth accentuating again that
though it is perfectly all right to analyze the data in the AB
scheme, one has to keep in mind that not all hard effects are
absorbed in the gluonic coefficient functions in such a scheme.

The sea-quark and gluon spin distributions cannot be separately
determined from current experimental data. In other words, while
the shapes of the spin-dependent valence quark distributions are
fairly constrained by the data, sea-quark and gluon spin densities
are almost completely undetermined. In particular, $\Delta G$ is
rather weakly constrained by the data. This is understandable
because the gluon polarization contributes to $g_1$ only at the
NLO level through the convolution $\Delta C_G\otimes\Delta G$. In
principle, measurements of scaling violation in $g_1(x,Q^2)$ via,
for example, the derivative of $g_1(x,Q^2)$ with respect to $Q^2$,
in next-generation experiments will allow an estimate of the gluon
spin density and the overall size of gluon polarization. Of
course, the data should be sufficiently accurate in order to study
the gluon spin density. Meanwhile, it is even more important to
probe $\Delta G(x)$ independently in those hadron-hadron collision
processes where gluons play a dominant role.

Though the polarized structure function is factorization scheme
independent, it is important to perform NLO analyses in different
schemes to test the reliability and consistency of the theory. It
is found in \cite{SMC98,Leader98pr,Leader99} that the polarized
parton distributions obtained in $\MS$, AB and CI schemes agree
well for the non-singlet spin densities, while the first moment of
$\Delta G(x)$ obtained in the AB or CI scheme is different from
that in the $\MS$ scheme, reflecting that the present data can
hardly constrain the gluon spin distribution. Typically, the
extracted value of the gluon spin at $Q^2=1\,{\rm GeV}^2$ lies in
the range $0\lsim\Delta G\lsim 2$\,. However, the recent analysis
\cite{Ghosh} shows that the LO fit cannot decide on the sign of
$\Delta G$, while the NLO analysis yields a negative first moment
of the gluon density.\footnote{It has been found by Jaffe
\cite{Jaffe96b} that the gluon spin component is negative $\Delta
G\sim -0.4$ in the MIT bag model and even more negative in the
non-relativistic quark model. However, it was explained in
\cite{Barone} that the negative $\Delta G$ obtained by Jaffe is a
consequence of neglecting some self-interaction effects.} This
illustrates again that it is difficult to pin down the gluon spin
distribution from present polarized DIS data. Note that a recent
NLO analysis in \cite{Leader99} shows that the polarized strange
quark density is significantly different from zero independently
of the factorization schemes used in the analysis:
\be
\Delta s_\MS=-0.102\pm 0.012,\quad \Delta s_{\rm CI}=-0.064\pm
0.010, \quad \Delta s_{\rm AB}=-0.058\pm 0.012\,,
\en
at $Q^2=1\,{\rm GeV}^2$. Note that the sea polarization $\Delta s$
is scheme dependent:
\be
\Delta s_{\rm CI,AB}=\Delta s_{\rm GI}+{\alpha_s\over
2\pi}\,\Delta G.
\en
The presence of the sea polarization in the CI or AB scheme
implies that the gluon polarization is not at its maximal value
given by $\Delta G\sim -(2\pi/\alpha_s)\Delta s_{\rm GI}$.

\subsection{Orbital angular momentum}
The orbital angular momentum plays a role in the proton spin
structure. For example, the growth of large $\Delta G$ with $Q^2$
is balanced by the large negative orbital angular momentum of the
quark-gluon pair. It is also known that the reduction of the total
spin component $\Delta\Sigma$ due to the presence of the quark
transverse momentum in the lower component of the Dirac spinor is
traded with the quark orbital angular momentum.

In the proton spin sum rule:
\be
\label{spinsr} {1\over
2}=J_q+J_G,
\en
the total angular momenta $J_q$ and $J_G$ of quarks and gluons
respectively are gauge invariant. However, the decomposition into
spin and orbital components, $J_q={1\over 2}\Delta\Sigma+L_q$ and
$J_G=\Delta G+L_G$, is gauge dependent. This leads to difficulties
in defining the partonic spin densities: There exist no local
gauge invariant operators that could represent the densities of
gluon spin and the orbital angular momenta of quarks and gluons.
It is known that the spin and orbital angular momenta of quarks
and gluons  appearing in the decomposition \cite{JM}
\be
J^z &=& J_q^z+J_G^z=S_q^z+L_q^z+S_G^z+L_G^z \non \\ &=& \int
d^3x\left[ {1\over
2}\bar\psi\gamma^3\gamma^5\psi+\psi^\dagger(\vec{x}\times
i\vec{\partial})^3\psi+(\vec{E}\times \vec{A})^3-E^k(\vec{x}\times
\vec{\partial})^3A^k\right],
\en
obtained in the the light-cone gauge $A^+=0$ and in the infinite
momentum frame, are separately gauge variant except for the quark
spin operator $S_q^z$. However, the gluon spin and its
distribution, for example, are physical, gauge invariant
quantities and can be measured experimentally. The point is that
$\Delta G$ can be expressed as the matrix element of a string-like
nonlocal gauge-invariant operator $O_G$ \cite{JM,BB}. As stressed
in \cite{Ji99}, what one measures experimentally is the matrix
element of the gauge-invariant operator $O_G$. But in the
light-cone gauge, the above operator reduces to the gluon spin
operator $S_G^z=\int d^3x(\vec{E}\times\vec{A})^3$; that is, the
gauge-invariant extension of the gluon spin operator in the
light-cone gauge is measurable. Likewise, a gauge-invariant
operator that reduces to the quark orbital angular momentum in the
light-cone gauge has been discussed in \cite{Jaffe98} (see however
a different discussion in \cite{Shore99}).

In principle, the total angular momenta of quarks and gluons,
$J_q$ and $J_G$ respectively, can be measured in deeply virtual
Compton scattering in a special kinematic region where single
quark scattering dominates \cite{Ji97}. It has also been suggested
that the orbital angular momentum might be deduced from an
azimuthal asymmetry in hadron production with a transversely
polarized target.

The evolution of the quark and gluon orbital angular momenta was
first discussed by Ratcliffe \cite{Rat87}.  Ji, Tang and Hoodbhoy
\cite{Ji1} have derived a complete leading-log evolution equation
in the light-cone gauge:
\be
{d\over dt}\left(\matrix{ L_q  \cr L_G
\cr}\right)={\alpha_s(t)\over 2\pi} \left(\matrix{ -{4\over 3}C_F
& {n_f\over 3} \cr {4\over 3}C_F & -{n_f\over 3}
\cr}\right)\left(\matrix{ L_q  \cr L_G
\cr}\right)+{\alpha_s(t)\over 2\pi} \left(\matrix{ -{2\over 3}C_F
& {n_f\over 3} \cr -{5\over 6}C_F & -{11\over 2}
\cr}\right)\left(\matrix{ \Delta\Sigma \cr \Delta G \cr}\right),
\en
with the solutions
\be \label{Lqg} && L_q(Q^2) = -{1\over
2}\Delta\Sigma+{1\over 2}\,{3n_f\over 16+3n_f}+
f(Q^2)\left(L_q(Q^2_0)+{1\over 2}\Delta\Sigma -{1\over
2}\,{3n_f\over 16+3n_f}\right),   \non \\ && L_G(Q^2) = -\Delta
G(Q^2)+{1\over 2}\,{16\over 16+3n_f}+
f(Q^2)\left(L_G(Q^2_0)+\Delta G(Q^2_0)-{1\over 2}\, {16\over
16+3n_f}\right), \non \\ &&
\en
where
\be
f(Q^2)=\left({\ln Q^2_0/\Lambda^2_{_{\rm QCD}}\over \ln
Q^2/\Lambda^2_{_{\rm QCD}} }\right)^{{32+6n_f\over 33-2n_f}}
\en
and $\Delta\Sigma$ is $Q^2$ independent to the leading-log
approximation. We see that the growth of $\Delta G$ with $Q^2$ is
compensated by the gluon orbital angular momentum, which also
increases like $\ln Q^2$ but with opposite sign. The solution
(\ref{Lqg}) has an interesting implication in the asymptotic limit
$Q^2\to \infty$, namely
\be
&& J_q(Q^2) = {1\over 2}\Delta\Sigma+L_q(Q^2)\to~~~ {1\over
2}\,{3n_f\over 16+3n_f}\,,    \non \\ && J_G(Q^2)=\Delta
G(Q^2)+L_G(Q^2)\to ~~~{1\over 2}\,{16\over 16+3n_f}\,.
\en
Thus, history repeats herself: The partition of the nucleon spin
between quarks and gluons follows the well-known partition of the
nucleon momentum. Taking $n_f=6$, we see that
$J_q:J_G=0.53\,:\,0.47\,$. If the evolution of $J_q$ and $J_G$ is
very slow, which is empirically known to be true for the momentum
sum rule that half of the proton's momentum is carried by gluons
even at a moderate $Q^2$, then $\Delta\Sigma\sim 0.25$ at
$Q^2=10\,{\rm GeV}^2$ implies that $L_q\sim 0.13$ at the same
$Q^2$, recalling that the quark orbital angular momentum is
expected to be of order 0.20 in the relativistic quark model.

Recently the quark orbital angular momentum of the nucleon was
calculated from lattice QCD by considering the form factor of the
quark energy-momentum tensor $T_{\mu\nu}$ \cite{Liu99}. The  total
angular momentum of the quarks is found to be
\be J_q=0.30\pm
0.07\,.
\en
That is about 60\% of the proton spin is attributable to the
quarks. Hence, the quark orbital angular momentum is $L_q=0.17\pm
0.06$ when combining with the previous quark spin content ${1\over
2}\Delta\Sigma=0.13\pm 0.06$ \cite{Dong}. Therefore, about 25\% of
the proton spin originates from the quark spin and about 35\%
comes from the quark orbital angular momentum.

It must be stressed that $J_q$ should be factorization scheme
independent. This means that a replacement of $\Delta\Sigma_{\rm
GI}$ by $\Delta\Sigma_{\rm CI}$ in the spin sum rule
(\ref{spinsr}) requires that the difference $\Delta\Sigma_{\rm
CI}-\Delta \Sigma_{\rm GI}=(n_f\alpha_s/2\pi)\Delta G$ be
compensated by a counterpart in the gluon orbital angular
momentum:
\be
L_q^{\rm CI}=L_q^{\rm GI}-{n_f\alpha_s\over 4\pi}\Delta G.
\en
It is interesting to note that if $\Delta G$ is of order 2.5\,,
one will have $\Delta\Sigma_{\rm CI}\sim 0.58$ and $L_q^{\rm
CI}\sim 0$\,. In other words, while $\Delta\Sigma_{\rm CI}$ is
close to the relativistic quark model value of $\Delta\Sigma$,
$L_q^{\rm CI}$ deviates farther from the quark model expectation
$L'_q\sim 0.20$ (see Sec. I).

\section{Conclusions}
\setcounter{equation}{0} The spin sum rule of the proton in the
infinite momentum frame reads
\be
{1\over 2}=J_q+J_G={1\over 2}\Delta\Sigma+L_q+\Delta G+L_G.
\en
The quark spin can be inferred from polarized DIS measurements of
$g_1(x)$ and its first moment. Due to the ambiguity arising from
the axial anomaly, the definition of the sea polarization $\Delta
q_s$ and hence $\Delta q=\Delta q_v+\Delta q_s$ is $k_\perp$
factorization dependent, but $J_q,~g_1(x)$ and $\Gamma_1$ are not.
The only spin content which is for sure at present is the observed
value $\Delta\Sigma\sim 0.20-0.30$ in the GI scheme (e.g. the
$\MS$ scheme). The recent lattice calculation yields $J_q=0.30\pm
0.07$. Therefore, we have $L_q^{\rm GI}\sim 0.10\pm 0.06$ for
$\Delta\Sigma_{\rm GI}\sim 0.25\,$, to be compared with the quark
model prediction $L_q'\sim 0.20$\,. The values of $\Delta\Sigma$
and $L_q$ in the CI scheme (e.g. the improved QCD parton model) or
the AB scheme depend on the gluon spin. Since $L_q^{\rm
CI}-L_q^{\rm GI}=-n_f\alpha_s\Delta G/(4\pi)$, it is clear that if
the gluon polarization is positive and large enough, then
$L_q^{\rm CI}$ will deviate even farther from the quark picture
although $\Delta\Sigma_{\rm CI}$ can be made to be close to the
constituent relativistic quark model. In the asymptotic limit,
$J_q(\infty)={1\over 2}\Delta\Sigma(\infty)+L_q(\infty)\sim{1\over
4}$ and $J_G(\infty) =\Delta G(\infty)+L_G(\infty)\sim {1\over
4}$. The recent lattice result $J_q=0.30\pm 0.07$ at a moderate
$Q^2$ seems to suggest that the evolution of $J_q$ and hence $J_G$
is slow enough.

\vskip 3 cm \centerline{\bf ACKNOWLEDGMENTS} \vskip 0.3 cm This
work was supported in part by the National Science Council of ROC
under Contract No. NSC89-2112-M001-016.

\renewcommand{\baselinestretch}{1.1}
\newcommand{\bi}{\bibitem}


\newpage
\begin{thebibliography}{99}
%

\bi{PDG} Particle Data Group, {\sl Eur. J. Phys.} {\bf C3}, 1
(1998).

\bi{Goto} Y. Goto {\it et al.,} hep-ph/0001046.

\bi{EJ} J. Ellis and R.L. Jaffe, \pr {\bf D9}, 1444 (1974); {\bf
D10}, 1669 (1974).

\bi{EMC} EMC, J. Ashman {\it et al.,} \np {\bf B238}, 1 (1990);
\pl {\bf B206}, 364 (1988).

\bi{Anselmino} M. Anselmino, A. Efremov, and E. Leader, {\sl Phys.
Rep.} {\bf 261}, 1 (1995).

\bi{Jaffe96a} R.L. Jaffe, MIT-CTP-2518 [hep-ph/9603422];
MIT-CTP-2506 [hep-ph/9602236].

\bi{EK96} J. Ellis and M. Karliner, CERN-TH/95-334
[hep-ph/9601280].

\bi{Cheng96b} H.Y. Cheng, {\sl In. J. Mod. Phys.} {\bf A11}, 5109
(1996).

\bi{Lampe} B. Lampe and E. Reya, hep-ph/9810270.

\bi{Shore98} G.M. Shore, hep-ph/9812355.

\bi{Bass99} S.D. Bass, hep-ph/9902280.

\bi{Schreiber} A.W. Schreiber and A.W. Thomas, \pl {\bf B215}, 141
(1988).

\bi{E80} SLAC-E80 Collaboration, M.J. Alguard {\it et al.,} \prl
{\bf 37}, 1261 (1976); \prl {\bf 41}, 70 (1978); G. Baum {\it et
al.,} \prl {\bf 45}, 2000 (1980).

\bi{E130} SLAC-E130 Collaboration, G. Baum {\it et al.,} \prl {\bf
51}, 1135 (1983).

\bi{E142} SLAC-E142 Collaboration, P.L. Anthony {\it et al.,} \prl
{\bf 71}, 959 (1993); \pr {\bf D54}, 6620 (1996).

\bi{E143} SLAC-E143 Collaboration, K. Abe {\it et al.,} \pr {\bf
D58}, 112003 (1998); \prl {\bf 74}, 346 (1995); ibid. {\bf 75}, 25
(1995).

\bi{E154} SLAC-E154 Collaboration, K. Abe {\it et al.,}  {\sl
Phys. Lett.} {\bf B405}, 180 (1997).

\bi{E155} SLAC-E155 Collaboration, P.L. Anthony {\it et al.,} \pl
{\bf B463}, 339 (1999).

\bibitem{SMC98} SMC, B. Adeva {\it et al.,} \pr {\bf D58}, 112001,
112002 (1998).

\bi{HERMES} HERMES Collaboration, K. Ackerstaff {\it et al.,} \pl
{\bf B404}, 383 (1997); A. Airapetian {\it et al.,} \pl {\bf
B442}, 484 (1998).

\bi{Lar91} S.A. Larin and J.A.M. Vermaseren, \pl {\bf B259}, 345
(1991).

\bi{Frank} L.L. Frankfurt, M.I. Strikman, L. Mankiewicz, A.
Sch\"afer, E. Rondio, A. Sandacz, and V. Papavassiliou, \pl {\bf
B230}, 141 (1989).

\bi{HERMESsemi} HERMES Collaboration, K. Ackerstaff {\it et al.,}
\pl {\bf B464}, 123 (1999).

\bibitem{SMCsemi} SMC, B. Adeva {\it et al.,} \pl  {\bf B420}, 180
(1998).

\bi{ChengLi96} T.P. Cheng and L.F. Li, \pl {\bf B366}, 365 (1996);
hep-ph/9811279.

\bi{HERMESdG} HERMES Collaboration, A. Airapetian {\it et al.,}
hep-ex/9907020.

\bi{Efr}  A.V. Efremov and O.V. Teryaev, JINR Report E2-88-287
(1988), and in {\it Proceedings of the International Hadron
Symposium}, Bechyne, Czechoslovakia, 1988, eds. Fischer {\it et
al.} (Czechoslovakian Academy of Science, Prague, 1989), p.302.

\bi{AR} G. Altarelli and G.G. Ross, \pl {\bf B212}, 391 (1988).

\bi{CCM} R.D. Carlitz, J.C. Collins, and A.H. Mueller, \pl {\bf
B214}, 229 (1988).

\bi{Leader88} E. Leader and M. Anselmino, Santa Barbara Preprint
NSF-ITP-88-142 (1988).

\bi{Kodaira} J. Kodaira {\it et al.,} \np {\bf B165}, 129 (1980);
ibid. {\bf B159}, 99 (1979).

\bi{Lam} C.S. Lam and B.A. Li, \pr {\bf D25}, 683 (1982).

\bi{Rat} P. Ratcliffe, \np {\bf B223}, 45 (1983).

\bi{JM} R.L. Jaffe and A.V. Manohar, \np {\bf B337}, 509 (1990).

\bi{BQ} G.T. Bodwin and J. Qiu, \pr {\bf D41}, 2755 (1990), and in
{\it Proc. Polarized Collider Workshop}, University Park, PA,
1990, eds. J. Collins {\it et al.} (AIP, New York, 1991), p.285.

\bi{Manohar} A.V. Manohar, in {\it Proc. Polarized Collider
Workshop}, University Park, PA, 1990, eds. J. Collins {\it et al.}
(AIP, New York, 1991), p.90; \prl {\bf 66}, 1663 (1991).

\bi{Leader98pl} E. Leader, A.V. Sidorov, and D.B. Stamenov, \pl
{\bf B445}, 232 (1998).

\bi{Bass} S.D. Bass, \zp {\bf C55}, 653 (1992).

\bi{Cheng96a} H.Y. Cheng, {\sl Chin. J. Phys.} {\bf 34}, 738
(1996) [hep-ph/9510280].

\bi{Mank} L. Mankiewicz, \pr {\bf D43}, 64 (1991).

\bi{Ball} R.D. Ball, S. Forte, and G. Ridolfi, \pl {\bf B378}, 255
(1996); R.D. Ball, hep-ph/9511330; S. Forte, hep-ph/9511345.

\bi{Cheng98} H.Y. Cheng, \pl {\bf B427}, 371 (1998).

\bi{Cheng97} H.Y. Cheng, {\sl Chin. J. Phys.} {\bf 35}, 25 (1997)
[hep-ph/9512267].

\bi{Muller} D. M\"uller and O.V. Teryaev, \pr {\bf D56}, 2607
(1997).

\bi{Smith} C.H. Llewellyn Smith, hep-ph/9812301.

\bi{Brod88} S.J. Brodsky, J. Ellis, and M. Karliner, \pl {\bf
B206}, 309 (1988).

\bi{Dor} A.E. Dorokhov and N.I. Kochelev, {\sl Mod. Phys. Lett.}
{\bf A5}, 55 (1990); \pl {\bf B245}, 609 (1990); \pl {\bf B259},
335 (1991); A.E. Dorokhov, N.I. Kochelev, and Yu.A. Zubov, {\sl
Int. J. Mod. Phys.} {\bf A8}, 603 (1993).

\bi{Dong} S.J. Dong, J.-F. Laga\"e, and K.F. Liu, \prl {\bf 75},
2096 (1995).

\bi{Fukugita} M. Fukugita, Y. Kuramashi, M. Okawa, and A. Ukawa,
\prl {\bf 75}, 2092 (1995).

\bi{Gockeler1} M. G\"ockeler, R. Horsley, E.-M. Ilgenfritz, H.
Perlt, P. Rakow, G. Schierholz, and A. Schiller, \pr {\bf D53},
2317 (1996).

\bi{Gockeler2} M. G\"ockeler, R. Horsley, E.-M. Ilgenfritz, H.
Perlt, P. Rakow, G. Schierholz, and A. Schiller, \pl {\bf B414},
340 (1997); S. Capitani {\it et al.,} hep-ph/9905573.

\bi{Gusken} S. G\"usken {\it et al.,} \pr {\bf D59}, 114502
(1999).

\bi{Zijl} E.B. Zijlstra and W.L. van Neerven, \np {\bf B417}, 61
(1994); {\bf B426}, 245(E) (1994).

\bi{Mert} R. Mertig and W.L. van Neerven, \zp {\bf C70}, 637
(1996); W. Vogelsang, \pr {\bf D54}, 2023 (1996); hep-ph/9603366,
hep-ph/9607223.

\bi{Gluck96} M. Gl\"uck, E. Reya, M. Stratmann, and W. Vogelsang,
\pr {\bf D53}, 4775 (1996).

\bi{Gehr96} T. Gehrmann and W.J. Stirling, \pr {\bf D53}, 6100
(1996).

\bi{SMC97} SMC, D. Adams {\it et al.,} \pl {\bf B396}, 338 (1997);
\pr {\bf D56}, 5330 (1997); B. Adeva {\it et al.,} \pl {\bf B412},
414 (1997).

\bibitem{Leader98pr} E. Leader, A. V. Sidrov, and D.B. Stamenov,
\pr {\bf D58}, 114028 (1998).

\bi{Florian98} D. de Florian, O. Sampayo, and R. Sassot, {\sl
Phys. Rev.} {\bf D57}, 5803 (1998).

\bibitem{Gordon} L.E. Gordon, M. Goshtasbpour, and G.P. Ramsey,
\pr {\bf D58}, 094017 (1998).

\bibitem{Bourrely98} C. Bourrely, F. Buccella, O. Pisanti, P. Santorelli,
and J. Soffer, {\sl Prog. Theor. Phys.} {\bf 99}, 1017 (1998).

\bi{Leader99} E. Leader, A.V. Sidorov, and D.B. Stamenov, \pl {\bf
B462}, 189 (1999).

\bi{Tatur} S. Tatur, J. Bartelski, and M. Kurzela, hep-ph/9903411.

\bi{Ghosh} D.K. Ghosh, S. Gupta, and D. Indumathi, hep-ph/0001287.

\bi{Altarelli97} G. Altarelli, R.D. Ball, S. Forte, and G.
Ridolfi, \np {\bf B496}, 337 (1997); {\sl Acta Phys. Pol.} {\bf
B29}, 1145 (1998).

\bi{Jaffe96b} R.L. Jaffe, \pl {\bf B365}, 359 (1996).

\bi{Barone} V. Barone, T. Calarco, and A. Drago, \pl {\bf B431},
405 (1998).

\bi{BB} I.I. Balitsky and V.M. Braun, \pl {\bf B267}, 405 (1991).

\bi{Ji99} P. Hoodbhoy, X. Ji, and W. Lu, \pr {\bf D59}, 074010
(1999).

\bi{Jaffe98} S.V. Bashinskii and R.L. Jaffe, \np {\bf B536}, 303
(1998).

\bi{Shore99} G.M. Shore and B.E. White, hep-ph/9912341.

\bi{Ji97} X. Ji, \prl {\bf 78}, 610 (1997).

\bi{Rat87} P.G. Ratcliffe, \pl {\bf B192}, 180 (1987).

\bi{Ji1} X. Ji, J. Tang, and P. Hoodbhoy, \prl {\bf 76}, 740
(1996).

\bi{Liu99} N. Mathur, S.J. Dong, K.F. Liu, L. Mankiewicz, and N.C.
Mukhopadhyay, hep-ph/9912289.

\end{thebibliography}
\end{document}